# Induced magneto-transport properties at palladium/yttrium iron garnet interface

Tao Lin, Chi Tang, and Jing Shi

Department of Physics and Astronomy, University of California, Riverside, CA 92521

As a thin layer of palladium (Pd) is directly deposited on an yttrium iron garnet or YIG ($Y_3Fe_5O_{12}$) magnetic insulator film, Pd develops both low- and high-field magneto-transport effects that are absent in standalone Pd or thick Pd on YIG. While the low-field magnetoresistance peak of Pd tracks the coercive field of the YIG film, the much larger high-field magnetoresistance and the Hall effect do not show any obvious relationship with the bulk YIG magnetization. The distinct high-field magneto-transport effects in Pd are shown to be caused by interfacial local moments in Pd.

Noble metals such as Pt and Au are preferred spin current generators or detectors [1-5] due to their strong spin-orbit interaction that results in a large spin Hall angle. The large room-temperature spin Hall effect quantified by the spin Hall angle plays an important role in spintronics. Among several interesting spin current related effects, the spin Seebeck effect (SSE) reported in magnetic metal [6], semiconductor [7], and insulator [8, 9] has received much attention. Recently, SSE in insulators in particular was challenged by a possible magnetic proximity effect (MPE) [10] existing at the Pt/magnetic insulator interface. It was shown that the MPE, along with the anomalous Nernst effect in the magnetized Pt interface layer can generate a significant SSE-like signal. More experiments have been carried out in different geometries and with a Au detector that is less prone to MPE [11]. At the heart of the debate, one key question is whether MPE exists in spin current detector materials where the inverse spin Hall effect is used.

Motivated by these experiments, we have chosen a different material, Pd, for this study. First of all, Pd is a 4d transition metal and has large magnetic susceptibility which favors MPE. It was shown that MPE does exist at Pd/ferromagnetic metal interfaces [12]. In addition, Pd also has strong spin-orbit interaction and been shown to have a large spin Hall conductivity [13]. To address the issue of possible MPE in magnetic insulator-based structures, we choose Pd/yttrium iron garnet (YIG) as our main material system, although we have also prepared Pt/YIG for comparisons. In this work, we focus on Pd/YIG samples.

Thin YIG films are grown on single crystal gallium gadolinium garnet (GGG) substrates with both (110) and (111) orientations using a pulsed laser deposition system. The base pressure of the deposition chamber is $6 \times 10^{-7}$ Torr. During growth, the chamber is back filled with ozone (1.5 mTorr) and the growth temperature is kept at ~ 700°C. Epitaxial YIG films are obtained as indicated by the reflection high-energy electron diffraction (RHEED) pattern shown in Fig. 1a. For this work, YIG films with thickness ranging from 100 to 200 nm are used. Both orientations show well-defined in-plane magnetic anisotropy, indicating dominance of the shape anisotropy. Because of the similarity in both orientations, this work only includes films in (111) orientation.

Typical magnetic hysteresis loops are shown in Fig. 1a. The out-of-plane saturation field is ~ 2 kOe, which corresponds well to $4\pi M_s$=1780 G for YIG. After YIG films are taken out of the PLD chamber, they are immediately placed in a high-vacuum sputtering chamber where a thin Pd film is deposited. Before deposition, YIG film is lightly sputtered to provide a fresh and clean surface. In this work, we focus on Pd films with a thickness range from 1.5 to 10 nm. Hall bars with the width of 200 μm and length of 1000 μm are patterned using ion milling. The magnetic properties of YIG films are measured with either a vibrating sample magnetometer or Quantum Design's magnetic property measurement system. The magneto-transport measurements are conducted in either a close-cycled refrigerator with an electromagnet (<1 T) or Quantum Design's physical property measurement system (up to 14 T).

As an in-plane magnetic field $H_{/\!/}$ is swept along the Hall bar direction, the magnetoresistance (MR), $\frac{\Delta\rho}{\rho} = \frac{\rho(H)-\rho(0)}{\rho(0)}$, of a Pd (2 nm)/YIG is shown in Fig. 2a, along with the magnetization data of YIG. Two negative peaks appear at the coercive fields of YIG. This feature resembles the anisotropic MR effect in ferromagnets. Here the MR peak is only ~6x10$^{-6}$, several orders smaller than that of the anisotropic MR in ferromagnetic conductors. MR with similar magnitude was previously reported in Pt/YIG where the YIG films are polycrystalline [10]. For comparison, a 2 nm thick Cu film deposited on YIG does not show any measurable MR signal. One possible cause of MR in Pd film is that the non-magnetic Pd film acquires a magnetic moment whose direction is dictated by the underlying YIG film, i.e. the Pd interface layer adjacent to YIG acting as if it is magnetic. As shown in Fig. 2b, the MR peaks are correlated with the coercive fields of YIG which do not change significantly with the temperature in this temperature range. However, the MR peak nearly doubles when the temperature is lowered to 30 K, which is consistent with reduced spin-flip scattering.

We have extended MR measurements to high fields. Fig. 3a shows MR of the same Pd/YIG sample with the field perpendicular to the film, $H_\perp$. Surprisingly, there is a much larger high-field magnetoresistance (HFMR) background that is overwhelmingly larger than the low-field MR signal shown in Fig. 2. At high temperatures, the positive MR is probably the usual Lorentz force induced effect. As the temperature is lowered, this positive MR diminishes and turns to negative. Negative MR is usually seen in materials with random spins that can be aligned by an external field to cause suppressed scattering. At the lowest temperature, the HFMR ratio reaches ~ -$10^{-3}$, nearly two orders greater than that of MR at low fields. The comparison between the low- and high-field MR reveals that in addition to the low-field phenomenon related to the YIG magnetization reversal, there is some spin-dependent process occurring at high fields. When additional spins are aligned with high fields, the MR ratio is consequently enhanced. It is interesting that the temperature dependence of HFMR (inset of Fig. 3a) is markedly different from that of the low-field MR.

In ferromagnetic conductors, superimposed on the ordinary Hall effect that is linear in $H_\perp$, there is a large anomalous Hall effect (AHE) signal that is proportional to the out-of-plane magnetization component [14]. However, in the low field range (up to ~2 kOe) where the in-plane magnetization is rotated towards the perpendicular direction and therefore there should be an AHE response, we do not observe any definitive magnetization-related AHE signal. As we ramp up $H_\perp$ further, however, an unambiguous non-linear AHE-like signal arises on the linear ordinary Hall background (removed in Fig. 3b). At low temperatures, there is a clear saturation in Hall resistivity at the highest magnetic field. The Hall resistivity reaches ~0.17 Ω at 5 K, equivalent to ~$1 \times 10^{-3}$ in the Hall angle. Note that the YIG magnetization saturates only with $H_\perp$ ~ 2 kOe, but saturation of the AHE-like signal does not occur until $H_\perp$ > 20 kOe. Therefore, similar to the HFMR effect, the high-field Hall signal also reveals a response of the magnetic moments other than those in the Pd interface layer that are possibly exchange aligned to the YIG magnetization. We fit the Brillouin function, i.e. $B_J(x) = \frac{2J+1}{2J}\coth(\frac{2J+1}{2J}x) - \frac{1}{2J}\coth(\frac{1}{2J}x); x = \frac{g\mu_B JB}{k_B T}$, to the AHE-like data in Fig. 3b. Here T is the temperature, $\mu_B$ is the Bohr magneton, and $gJ$ is treated as a fitting parameter. The solid curves in Fig. 3b are the actual Brillouin fits. Clearly, the saturation AHE-like signal steadily increases at low temperatures. The inset shows a plot of the normalized AHE-like signal as a function of B/T, indicating that the effective magnetic moment is not a temperature-independent constant. "$gJ$" decreases from ~ 200 $\mu_B$ at room temperature to ~ 7 $\mu_B$ at 5 K. It is known that a Fe impurity can induce a large local moment of in Pd [15, 16]; however, its temperature dependence has not yet been reported or understood.

Fig. 4a shows the Pd thickness dependence of the AHE-like signal in Pd/YIG samples. As the Pd thickness increases, the Hall magnitude sharply decreases. The inset shows the zoom-in plot of the AHE-like data for 4, 5, and 10 nm thick Pd films at room temperature. For 10 nm thick Pd, the Hall signal essentially vanishes. The rapidly decreasing trend of the AHE-like signal clearly demonstrates the interfacial origin of the magnetic moments that are responsible for the high-field Hall effect. Since the moments are located at the interface and it is the interface layer that produces a Hall signal, when the film thickness is much greater than the interface layer, the measured Hall voltage is quickly reduced due to the parallel resistance of the bulk Pd layer. For the same nominal 2 nm thick Pd on YIG, we have observed AHE with similar magnitude in five different samples.

Fig. 4b further reveals the properties of the interface moments. First of all, Pd needs to be in direct contact with YIG. Pd on MgO does not produce any Hall signal; therefore, the source of the interface moments must be YIG. Second, Cu either has no interface moments or does not produce any Hall signal even if it has interface moments. We cannot distinguish these two possibilities. If the latter is true, a 6 nm thick Cu layer is sufficiently thicker than the mean-free-path so that Pd does not feel any effect from the magnetic moments at the Cu/YIG interface. Third, interface roughness seems to enhance the Hall signal. The sputter cleaned YIG surface is likely rougher than the one without sputter cleaning and the Hall magnitude is a factor of 5 larger in the sample with a rough interface.

The above experimental facts strongly suggest that independent magnetic moments producing the high-field effects originate from the Pd/YIG interface. On the other hand, those moments are not exchange coupled to the YIG spins. In ferromagnetic conductors, the carriers are spin polarized and AHE arises from either extrinsic or intrinsic mechanisms due to spin-orbit interaction. But the existence of an AHE-like signal does not prove ferromagnetism. In the framework of AHE, the magnitude of AHE, $\rho_{xy}$, scales with the resistivity, $\rho_{xx}$, either linearly or quadratically, i.e. $\rho_{xy} \sim \rho_{xx}^n$, with n=1 or 2, depending on the microscopic mechanism [14]. In our Pd/YIG, the resistivity changes only ~ 18% but the AHE-like signal rises by a factor of 10 below 100 K. We do not expect any sharp temperature dependence of the saturation or fully aligned magnetic moments. Therefore, the dramatic rise of measured AHE-like signal at low temperatures argues against the AHE mechanism for spin-polarized carriers as in regular ferromagnets. Similar high-field effects were previously found in noble metal-based dilute magnetic alloys where the local moments can cause a left-right asymmetry to unpolarized electrons [17-19]. The Hall angle can be as large as $10^{-3}$ to $10^{-2}$. Either the spin-orbit interaction or spin-spin exchange between the local moments and the conduction electrons can result in such a Hall angle. The former is called the skew scattering [19] and the latter the "spin effect" [20]. The "spin effect" causes an enhanced ordinary Hall signal and MR, both of which vary with $<S_z>^2$, and therefore have a zero initial slope at H =0. This disagrees with our observations. Our experimental data in Pd/YIG are consistent with the skew scattering picture in which unpolarized electrons are deflected by local moments via spin-orbit interaction, similar to the noble metal-based dilute magnetic alloys [21]. We should point out that Pt/YIG samples also exhibit similar characteristic high-field features as observed in Pd/YIG but with larger magnitude in the Hall signals.

In summary, we have observed a low-field MR effect in Pd/YIG which tracks the bulk YIG magnetization reversal. In addition, we have also observed two different, much stronger magneto-transport effects that occur at high magnetic fields where the bulk YIG magnetization is already fully saturated. We attribute the observed Hall effect to the scattering of conduction electrons in Pd by local magnetic moments at Pd/YIG interface.

Acknowledgement: we thank F. Wang, Q. Niu, R.Q. Wu, and V. Aji for many enlightening discussions. TL and CT were supported by DMEA/CNN; JS was supported by a NSF/EECS grant.

Fig. 1. (a) Normalized magnetic hysteresis loops at 300 K of YIG film on GGG(111) substrate with an applied field in-plane ($H_\parallel$) and out-of-plane ($H_\perp$). Inset: RHEED pattern of YIG film on GGG(111). (b) Schematic diagram of the patterned Hall bar.

Fig. 2. (a) In-plane low-field MR of Pd(2 nm)/YIG (red squares), MR of Cu/YIG reference sample (black squares), and in-plane hysteresis loop (blue squares). (b) Temperature dependence of MR ratio of Pd(2 nm)/YIG.

Fig. 3. (a) Out-of-plane HFMR at different temperatures. The inset shows MR ratio at H=10 kOe. Red and blue regions represent positive and negative MR ratios respectively. (b) Field dependence of the Hall resistance $R_H$ at different temperatures with linear background removal. Lines are the Brillouin function fits. Inset: Normalized $R_H$ as a function of B/T shows that "gJ" changes as temperature is varied.

Fig. 4. (a) Pd thickness dependence of the Hall resistance $R_H$ at T=5 K. The inset shows the zoom-in data for Pd thicknesses from 4 to 10 nm. (b) $R_H$ for several reference samples at T=5 K. All metal layers (except the Cu-layer in Pd/Cu/YIG) are 2 nm thick. The inset shows zoom-in data for reference samples.

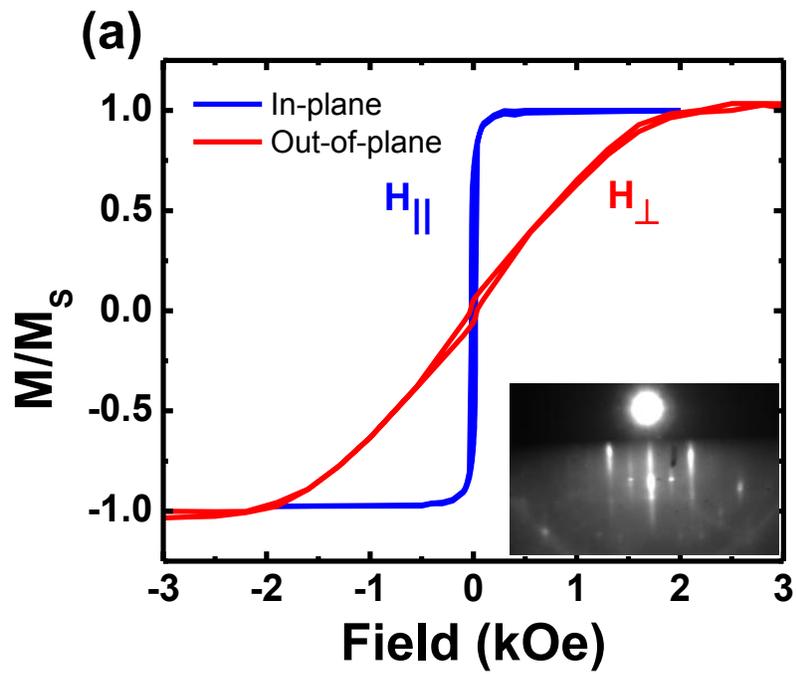

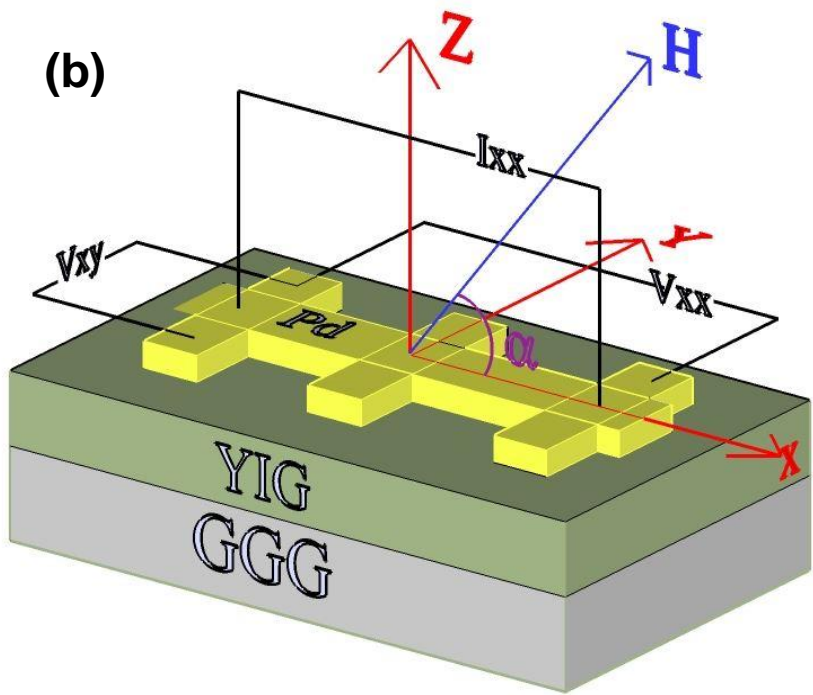

Figure 1

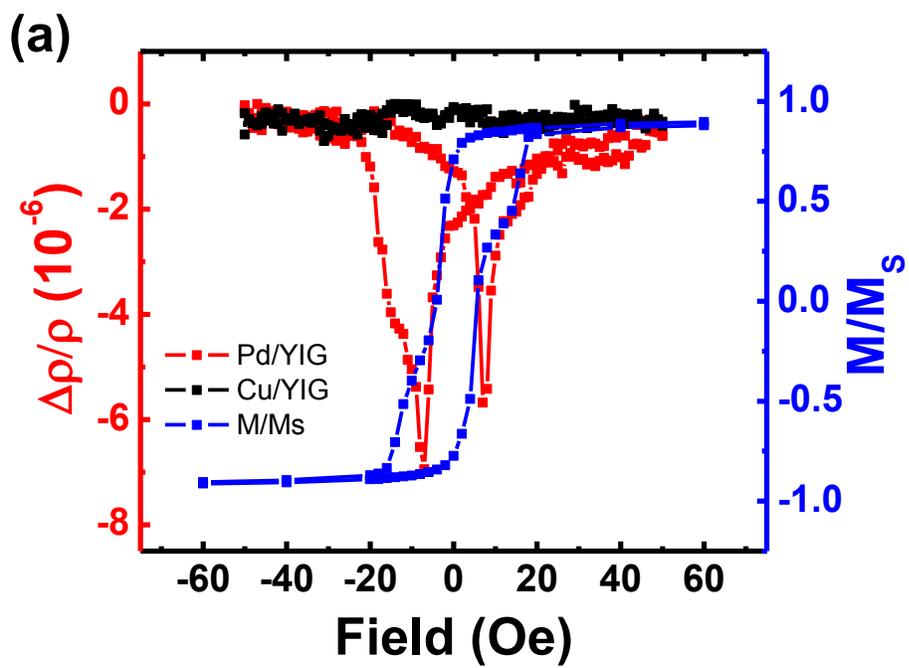

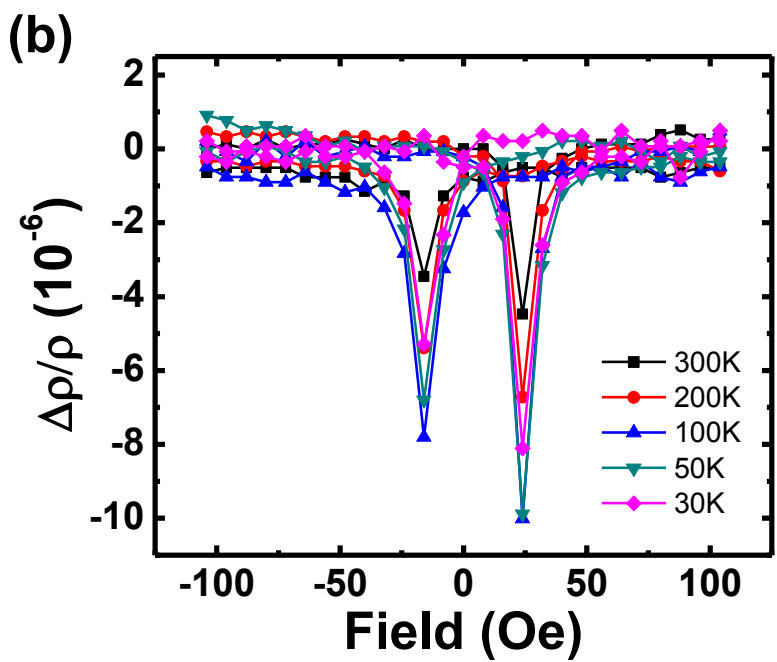

Figure 2

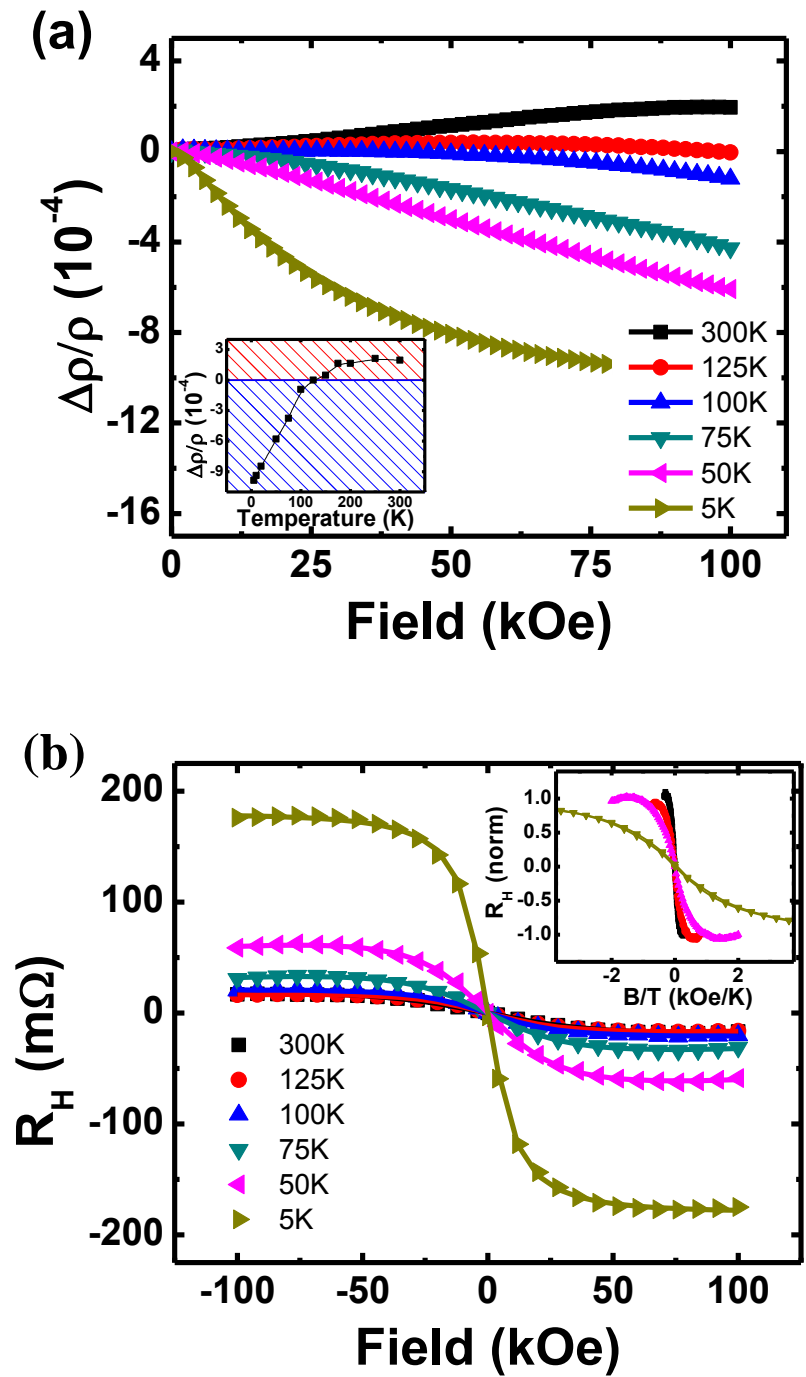

Figure 3

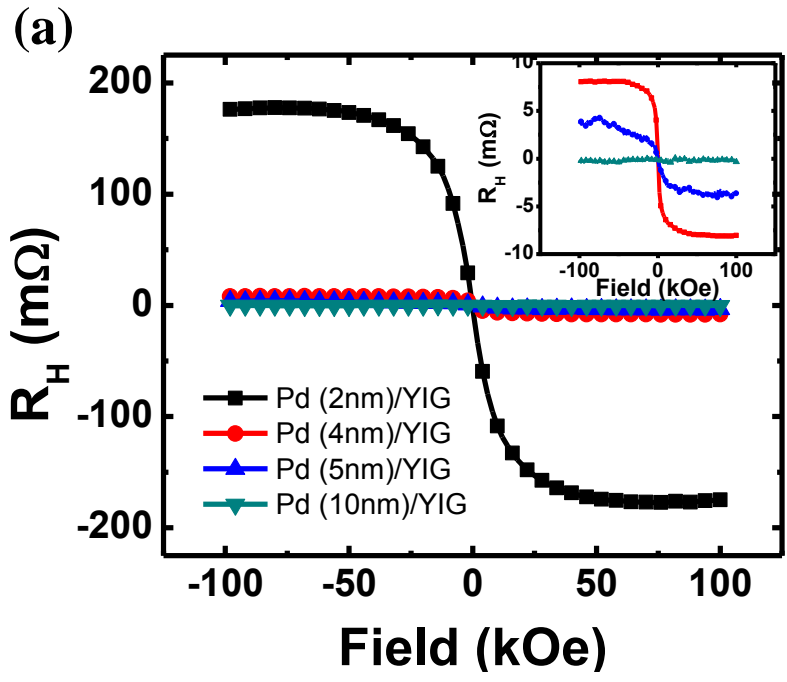

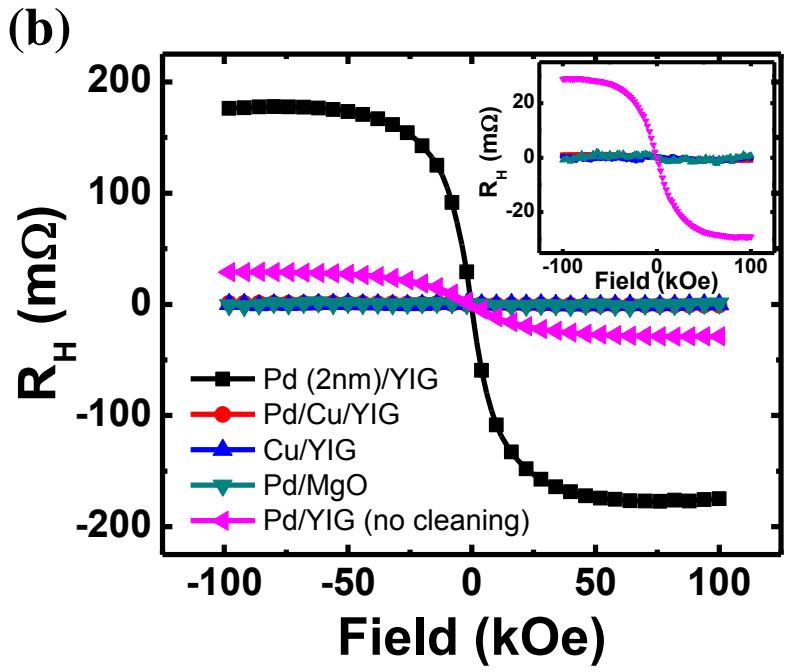

Figure 4

References:

[1]  E. Saito, M. Ueda, H. Miyajima, and G. Tatara, Appl. Phys. Lett. **88**, 182509 (2006).
[2]  T. Kimura, Y. Otani, T. Sato, S. Takahashi, and S. Maekawa, Phys. Rev. Lett. **98**, 156601 (2007).
[3]  T. Seki, Y. Hasegawa, S. Mitani, S. Takahashi, H. Imamura, S. Maekawa, J. Nitta, and K. Takanashi, Nature Mater. **7**, 125 (2008).
[4]  Y. Kajiwara, K. Harii, S. Takahashi, J. Ohe, K. Uchida, M. Mizuguchi, H. Umezawa, H. Kawai, K. Ando, K. Takahashi, S. Maekawa, and E. Saitoh, Nature (London) **464**, 262 (2010).
[5]  L. Liu, R. A. Buhrman, and D. C. Ralph, ArXiv e-prints, arXiv:1111.3702 (2011).
[6]  K. Uchida, S. Takahashi, K. Harii, J. Ieda, W. Koshibae, K. Ando, S. Maekawa, and E. Saitoh, Nature (London) **455**, 778 (2008).
[7]  C. M. Jaworski, J. Yang, S. Mack, D. D. Awschalom, J. P. Heremans, and R. C. Myers, Nature Mater. **9**, 898 (2010).
[8]  K. Uchida, J. Xiao, H. Adachi, J. Ohe, S. Takahashi, J. Ieda, T. Ota, Y. Kajiwara, H. Umezawa, H. Kawai, G. E. W. Bauer, S. Maekawa, and E. Saitoh, Nature Mater. **9**, 894 (2010).
[9]  K. Uchida, H. Adachi, T. Ota, H. Nakayama, S. Maekawa, and E. Saitoh, Appl. Phys. Lett. **97**, 172505 (2010).
[10] S. Y. Huang, X. Fan, D. Qu, Y. P. Chen, W. G. Wang, J. Wu, T. Y. Chen, J. Q. Xiao, and C. L. Chien, Phys. Rev. Lett. **109**, 107204 (2012).
[11] T. Kikkawa, K. Uchida, Y. Shiomi, Z. Qiu, D. Hou, D. Tian, H. Nakayama, X. –F. Jin, and E. Saitoh, Phys. Rev. Lett. **110**, 067207 (2013).
[12] J. Vogel, A. Fontaine, V. Cros, F. Petroff, J. Kappler, G. Krill, A. Rogalev, and J. Goulon, Phys. Rev. B **55**, 3663 (1997).
[13] G. Y. Guo, J. Appl. Phys. **105**, 07C701 (2009).
[14] N. Nagaosa, J. Sinova, S. Onoda, A. H. MacDonald, and N. P. Ong, Rev. Mod. Phys. **82**, 1539 (2010).
[15] A. J. Manuel, and M. McDougald, J. Phys. C **3**, 147 (1970).
[16] G. Bergmann, Phys. Rev. B **23**, 3805 (1981).
[17] A. Hamzic, S. Senoussi, I.A. Campbell, and A. Fert, Solid State Comm. **26**, 617 (1978).
[18] A. Hamzic, S. Senoussi, I.A. Campbell, and A. Fert, J. Magn. Magn. Mater. **15**, 921 (1980).
[19] A. Fert and O. Jaoul, Phys. Rev. Lett. **28**, 303 (1972).
[20] M.T. Beal-Monod, and R.A. Weiner, Phys. Rev. B **3**, 3056 (1971).
[21] A. Fert, A. Friederich, and A. Hamzic, J. Magn. Magn. Mater. **24**, 231 (1981).